\documentclass[12pt,osajnl,preprint,showpacs]{revtex4}
\usepackage{graphicx}

\begin{document}

\title{Dispersive  effects on optical information storage
in Bose-Einstein condensates using ultra-slow short pulses}

\author{Devrim Tarhan$^{1,2,3}$}
\author{Alphan Sennaroglu$^{2}$}
\author{\"{O}zg\"{u}r E. M\"{u}stecapl\i{}o\~{g}lu$^{2}$}

\affiliation{$^1$ Department of Physics, Faculty of Sciences and
Letters,\\
Istanbul Technical University, Maslak 34469, Istanbul, Turkey \\
$^2$ Department of Physics, Ko\c{c} University, Rumelifeneri yolu,
Sar\i{}yer, Istanbul, 34450, Turkey \\ $^3$ Department of Physics,
Harran University, Osmanbey Yerle\c{s}kesi, \c{S}anl\i{}urfa,
Turkey}

\email{dtarhan@ku.edu.tr}

\begin{abstract}

We investigate potential of atomic Bose-Einstein condensates as
dynamic memory devices for coherent optical information
processing. Specifically, the number of ultra-slow pulses that can
be simultaneously present within the storage time in the
condensate has been analyzed. By modelling short pulse propagation
through the condensate, taking into account high-order dispersive
properties, constraints on the information storage capacity have
been discussed. The roles of temperature, spatial inhomogeneity,
the interatomic interactions and the coupling laser on the pulse
shape have been pointed out. For a restricted set of parameters,
it has been found that coherent optical information storage
capacity would be optimized.

\end{abstract}


\maketitle
\section{Introduction}

The remarkable achievement of ultraslow
light\cite{slowlight-exp1,slowlight-exp2,slowlight-exp3} in a
Bose-Einstein condensate (BEC), using electromagnetically induced
transparency\cite{eit1,eit2,eit3} (EIT), inspired many intriguing
and attractive
applications\cite{slowlight-apps1,slowlight-apps2,slowlight-apps3,slowlight-apps4},
in particular quantum optical dynamic
memories\cite{qmemory1,qmemory2,qmemory3,qmemory4,qmemory5}. A key
component of a quantum information processor would be the dynamic
memory storage device. BECs have the potential to be used in such
an application. Consider a sequence of information bits carried by
a train of optical pulses. Upon injection into a BEC, the coherent
optical information can be stored for a period given by the
transit time of the pulses. In the BEC phase, the demonstrated
ultra-slow group velocities allow for long storage periods of the
order of few microseconds. Operational performance of such a
device can be characterized by the overall bit storage capacity of
the BEC, which can be defined as the number of optical pulses that
can be simultaneously stored during the storage time in the BEC.

In order to make quantum memories more practical for quantum
information processing, it is useful to enhance their bit storage
capacity. This would be possible, in principle, by using pulses
temporally as short as possible so that more pulses can be
injected into the BEC during the storage time\cite{liu}. Early
ultraslow light experiments use pulses of widths few microseconds.
Shorter pulses would have longer frequency bandwidths, beyond the
EIT window. EIT window can be modified by an external magnetic
field\cite{wei}. Very recently, it has been experimentally
demonstrated that EIT in a resonant atomic medium is possible for
a comb of picosecond pulses\cite{shortpulse-exp}. Promising
schemes to introduce large controllable time delays for such
ultrashort pulses has been proposed where EIT window is overcome
by using spatial and temporal processing\cite{shortpulse-prop}.
Distortions in the shape of a large-bandwidth pulse have been
experimentally studied in warm Rb vapor\cite{bashkansky}. Shapes
of the stored pulses will depend on the dispersive characteristics
of the BEC. In particular, each pulse in the train will undergo
dispersive broadening. The temporal separation of the pulses
should be adjusted in such a way that after broadening in the BEC,
consecutive pulses do not have significant spatial overlap. This
sets the limit to the bit storage capacity of the BEC and can be
estimated from the overall group delay and pulse broadening. It is
therefore necessary to investigate the dispersive effects inside a
BEC and develop accurate models that describe pulse propagation.

In this paper, we consider propagation of short pulses with widths
in the microseconds to nanoseconds range. In particular, we
investigate the role of dispersion on the temporal characteristics
of optical pulses propagating inside a BEC. Analytical formulas
are first summarized for the constitutive relations from which the
group delay, absorption coefficient, and the second-order group
delay dispersion can be calculated. We find that third and
higher-order dispersion effects are negligible. The influence of
the spatial inhomogeneity of the BEC, the interatomic scattering
interactions and the temperature on the pulse shape are taken into
account in our treatment. We show that just below the critical
temperature of condensation, broadening becomes maximum. Finally,
we give analytical formulas for the bit storage capacity of a BEC
and determine its optimum performance.

\section{Propagation and dispersion of slow-light pulses in
an atomic ensemble under EIT scheme}
\begin{figure}[htbp]
\centering
\includegraphics[width=8.3cm]{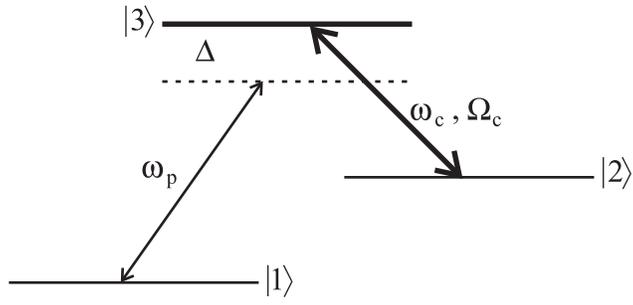}
\caption{Schematic energy level diagram of a
three-level atom interacting with two laser
beams in $\Lambda$ configuration. tarhanF1.eps.}
\label{fig1}
\end{figure}

We consider a gas of  $N$ three-level atoms interacting with two
laser beams in $\Lambda$ configuration as shown in Fig.\ref{fig1}.
In this scheme, it is assumed that the only dipole forbidden
transition is the one between the lower levels. The upper level is
coupled to the lower levels via a strong drive field with
frequency $\omega_c$ and a weak probe field of frequency
$\omega_p$. This particular arrangement leads to the well-known
effect of electromagnetically induced transparency (EIT) in which
the absorption of the probe field at resonance is cancelled. The
underlying physical mechanism of EIT is the quantum interference
induced by the drive field. Quantum coherence established between
the lower level doublet put them in a dark state where excitations
to the upper level become impossible. The transition probabilities
destructively interfere for the two possible excitation routes to
the upper level. In addition to turning an optically thick,
opaque, medium transparent, EIT has recently been used for
achieving ultra-slow light velocities, owing to the steep
dispersion of the EIT susceptibility near the probe resonance.
Susceptibility $\chi$ for the probe transition can be calculated
as a linear response as most of the atoms remain in the lowest
state. Assuming local density approximation, neglecting local
field, multiple scattering and quantum corrections and employing
steady state analysis, $\chi$ is found to be\cite{eit3}
\begin {equation}
\label{chieit}
\chi = \frac{\rho|\mu_{31}|^2}{\epsilon_0 \hbar}
\frac{i(i \Delta + \Gamma_2/2)}{[(\Gamma_2/2 + i
\Delta)(\Gamma_3/2 + i \Delta) + \Omega_c^2/4]},
\end {equation}
where $\Delta=\omega_{31}-\omega_p$ is the frequency detuning of
the probe field with frequency $\omega_p$ from the resonant
electronic transition $\omega_{31}$. For the cold gases considered
in this paper and assuming co-propagating laser beams, Doppler
shift in the detuning is neglected. It is assumed that the drive
field is at resonance. $\Omega_c$ is the Rabi frequency of the
drive field; $\mu_{31}$ is the dipole matrix element between
states $|3\rangle$ and $|1\rangle$ which can also be expressed in
terms of resonant wavelength $\lambda_{31}$ of the probe
transition via
$\mu_{31}^2=3\epsilon_0\hbar\lambda_{31}^3\gamma/8\pi^2$ with
$\gamma$ is the radiation decay rate between $|3\rangle$ and
$|1\rangle$. $\Gamma_2$ and $\Gamma_3$ denote the dephasing rates
of the atomic coherences between the appropriate states. $\rho$
stands for a given atomic density.

Probe beam propagates along the condensate axis in the $z$
direction. We treat the axial propagation of probe beam under
paraxial approximation where the paraxial effects and diffraction
losses are ignored. This approximation is valid when the probe
beam radius is much larger than the radial size of the atomic
cloud. In practice, using a pinhole and a flipper mirror, only
that portion of the probe beam passing through the thin central
column region of the cloud is selectively
monitored\cite{slowlight-exp1}. To develop a one dimensional wave
equation including the high order diffraction effects in the
dispersive medium for the short probe pulse, we follow
Ref.\cite{Harris}.

The probe pulses we shall consider in this paper have carrier
frequencies in the order of $10^{15}$ Hz, while their spectral
widths are less than of the order of $10^{9}$ Hz. We will study
their propagation under slowly varying phase and envelope
approximations. The wave equation that would govern the
propagation of a short pulse in a dispersive medium can be found
to be\cite{Harris}
\begin{equation} \label{eq:pulse}
\frac{\partial E}{\partial z} + \alpha E + \frac{1}{v_g}
\frac{\partial E}{\partial t} + i \,b_{2} \frac{\partial^2
E}{\partial t^2}  = 0,
\end{equation}
where $\alpha$ determines the pulse attenuation; $v_g$ is the
group velocity, and $b_{2}$ is the second-order dispersion
coefficient, or the group velocity dispersion. We have also
calculated the third order dispersive coefficient $b_3$ and found
it to be $7$ orders of magnitude less than $b_2$ for typical
experimental parameters\cite{slowlight-exp1}. The significant
coefficients for the short pulse propagation can be calculated
from the susceptibility using the relations\cite{Harris}
\begin{eqnarray}
\alpha &=& -\frac{i \pi}{\lambda} \chi(\omega_{0}),\\
\frac{1}{v_g} &=& \frac{1}{c} - \frac{\pi}{\lambda} \frac{\partial
\chi}{\partial \omega}|_{\omega_{0}},\label{eq:vg}\\
b_{2} &=& \frac{\pi}{2\lambda}
\left[\frac{\partial^{2}\chi}{\partial\omega^2}|_{\omega_{0}}\right].
\end{eqnarray}
Here we take $\lambda_{31}=\lambda$, $\omega_p\equiv\omega$ and
$\omega_{31}\equiv\omega_0$ for convenient notation. It may be
noted that these expressions lead to complex parameters in
general, as the $\chi$ is a complex valued function and the
parameters are merely mathematical coefficients in
Eq.\ref{eq:pulse}. At EIT resonance however they lead to
physically meaningful, well-defined, real valued absorption
coefficient, group velocity and dispersion coefficient. EIT
susceptibility exhibits steep normal dispersive behavior around
the resonance, which allows for the second term in Eq.\ref{eq:vg}
become much larger than the first term. This leads to substantial
reduction in the group velocity of the pulse. Such a slow pulse
propagates through the medium without much absorption due to the
small imaginary EIT susceptibility at resonance. On the other
hand, for short pulses the pulse shape would also be influenced by
the higher order dispersive coefficients, starting from the group
velocity dispersion. Different frequency components would have
different velocities and broadening of the pulse may become
significant. Let us emphasize that essential physical foundations
of these effects are not related to the density $\rho$ but the
dispersive properties of the EIT susceptibility. Indeed, slow
laser pulses have been observed in various media, including hot
rubidium vapor\cite{slowlight-hotgas1,slowlight-hotgas2},
ultracold Bose gas\cite{slowlight-exp1}, and in solid
crystals\cite{slowlight-solid}. When Rabi frequency for the
coupling field is sufficiently large such that $\Omega_c\gg
\Gamma_{2,3}$, we can calculate these parameters characterizing
propagation and dispersion of slow EIT pulses explicitly as
follows
\begin{eqnarray}
\alpha &=& \frac{2\pi\rho|\mu_{31}|^2\Gamma_2}
{\epsilon_0\hbar\lambda\Omega_c^2},\\
v_g &=&
\frac{c\epsilon_0\hbar\Omega_c^2}{2\omega_{31}|\mu_{31}|^2\rho},\label{vg}\\
b_2&=&i\frac{8\pi\Gamma_3|\mu_{31}|^2\rho}{\epsilon_0\hbar\lambda\Omega_c^4}.
\end{eqnarray}
It should be noted that the result for $v_g$ is valid when
\begin{eqnarray}
\Gamma_{2,3}\ll\Omega_c\ll\sqrt{4\pi c|\mu_{31}|^2\rho/\epsilon_0\hbar}
=\sqrt{\frac{3}{4\pi}c\lambda^3\gamma\rho}.
\end{eqnarray}
For $\rho\sim 10^{20}-10^{21}\,$m$^{-3}$, the upper limit would be
$5\gamma-15\gamma$. In this paper we consider
$\Omega_c=0.56\gamma-5\gamma$. Corresponding electric field of the
coupling laser then would be in the range $10^2-10^3\,$V/m, with
power densities $13-1300\,$ W/m$^2$. $\Omega_c=0.56\gamma$ is used
in typical experiments\cite{slowlight-exp1}. We shall discuss
possible enhancement of bit storage capacity by selecting
particular $\Omega_c$ within this range. According to their
dependencies on the $\Omega_c$, we see that increasing $\Omega_c$
reduces the dispersive effects significantly, but simultaneously
it reduces the storage time of the pulses in the condensate as the
group velocity would rise. It is illuminating to realize that
decrease of delay time (storage time) is much slower than the
decrease of dispersion coefficient with the $\Omega_c$. This can
be exploited to find a critical $\Omega_c$ for which number of
probe pulses injected into the medium within the storage time
would be optimized.

We note that for extremely large $\Omega_c$, one should use the following expression instead
\begin{eqnarray}
v_g=c\left(1-\frac{4\pi c|\mu_{31}|^2\rho}{\epsilon_0\hbar\Omega_c^2}\right)
\end{eqnarray}

For a uniform density medium the Eq.\ref{eq:pulse} can be solved
analytically\cite{bahaa}.
An initial temporally short gaussian
pulse after propagating in the medium of length $L$ is then found
to be delayed with respect to a reference pulse propagating in
vacuum by $t_d=L/v_g-L/c$. During that time, the pulse would
broaden due to second-order dispersion. Leaving the medium, it
would grow to a final width of
\begin{eqnarray}
\tau(L)=\tau_0\sqrt{1+(L/z_0)^2},
\label{eq:pulsewidth}
\end{eqnarray}
with $\tau_0$ is the initial temporal width of the pulse, and
$z_0=-\pi\tau_0^2/b_2$. For $L\gg z_0$ we get
$\tau(L)=|b_2|L/\pi\tau_0$.

In general, $\rho=\rho(r,z)$ is a spatially varying function,
whose profile depends on temperature, inter-atomic interactions
and confinement potential. Assuming slow spatial variations within
an optical wavelength the wave equation remains the same. The only
effect of spatial inhomogeneity would be to give group velocity
and its dispersion local character. As a result, the time delay
$t_d$ can be calculated through a spatial averaging of the group
velocity field\cite{morigi}. Experimentally measured group
velocity is then operationally defined by $v_g=L/t_d$, where the
effective axial length of the medium is evaluated by
\begin{eqnarray}
\label{eq:axial_size}
L=\left[\frac{4\pi}{N}\int_0^\infty\,r\mathrm{d}r\int_0^\infty\,\mathrm{d}z
z^2\rho(r,z)\right]^{1/2}.
\end{eqnarray}
The axial length $L$ is an effective length corresponding to the axial width of the density
distribution. It may be noted that more exact treatments can be used to
determine the effective length of an interacting BEC as discussed
in detail very recently\cite{length}. There is no simple
operational definition for the broadening of the pulse. One has to
either determine the output pulse shape from exact numerical
simulations of the wave equation or refer to uniform density
results for qualitative studies. While it is the atomic density
which determines the local group velocity and its dispersion,
after spatial averaging, the group velocity and broadening of the
pulse would become explicitly dependent on the temperature,
interactions between the atoms as well as the other physical
parameters due to the confinement potential. In order to elucidate
these dependencies, it is necessary to define $\rho$ concretely
and discuss how it is shaped by such physical parameters. We shall
specifically consider an atomic Bose-Einstein condensate as the
medium of propagation in this paper and review its density profile
in the next section.
\section{Density profile of a trapped semi-ideal Bose gas at finite temperature}
At low temperatures a Bose gas can be considered to be composed of
two components. The first component becomes the condensate part
and the other is a thermal gas background. In a harmonic trap,
spatial overlap between the components becomes small, which makes
effects of collective elementary excitations weaker relative to
those in homogeneous gases. This allows for semiclassical
descriptions of density distributions in terms of the
self-consistent Hartree-Fock model, which simplifies more general
Hartree-Fock-Popov description\cite{giorgini}. When the thermal
component is dilute enough, one may further neglect the atomic
interactions in the thermal component. Taking into account mean
field repulsion only among the condensed atoms, an analytical
explicit description of a partly condensed gas was developed and
denoted as the semi-ideal model\cite{naraschewski}. We will now
briefly review this model, where condensate density is evaluated
under Thomas-Fermi approximation\cite{tfa1,tfa2} to
Gross-Pitaevskii equation\cite{gpe1,gpe2}, and the thermal gas
density is calculated semi-classically\cite{Bagnato,naraschewski}.
The total density at a temperature $T$ is then written to be
\begin {equation}
\label{positiondependentro}
\rho(\vec{r}) = \frac{\mu-V(\vec{r})}{U_{0}}
\Theta(\mu-V(\vec{r})) \Theta(T_C-T) + \frac{g_{3/2} (z e^{-\beta
V})}{\lambda_T^3},
\end {equation}
where $U_0=4\pi\hbar^2 a_{s}/m$; $m$ is atomic mass; $a_s$ is the
atomic s-wave scattering length; $\mu$ is the chemical potential;
$\Theta(.)$ is the Heaviside step function;
$g_{n}(x)=\Sigma_{j}\,x^j/j^n$ is the Bose function; $\lambda_T$
is the thermal de Br\"{o}glie wavelength; $\beta=1/k_{B}T$;
$z=\exp{(\beta\mu)}$ is the fugacity, and $T_C$ is the critical
temperature. The external trapping potential is $V(\vec{r})=(m/2)
(\omega_r^2 r^2+\omega_z^2 z^2)$ with $\omega_r$ the radial trap
frequency and $\omega_z$ the angular frequency in the z direction.
$\mu$ is determined from
$N=\int\,\mathrm{d}^3\vec{r}\rho(\vec{r})$. At temperatures below
$T_c$ this yields\cite{naraschewski}
\begin{eqnarray}
\mu=\mu_{TF}\left(\frac{N_0}{N}\right)^{2/5},
\end{eqnarray}
where $\mu_{TF}$ is the chemical potential evaluated under
Thomas-Fermi approximation and the condensate fraction is given by
\begin{eqnarray}
\frac{N_0}{N}=1-x^3-s\frac{\zeta(2)}{\zeta(3)}x^2(1-x^3)^{2/5},
\end{eqnarray}
with $x=T/T_c$, and $\zeta$ is the Riemann-Zeta function. The
scaling parameter $s$, characterizing the strength of atomic
interactions within the condensate, is calculated to
be\cite{naraschewski,giorgini}
\begin{eqnarray}
s=\frac{\mu_{TF}}{k_BT_C}=\frac{1}{2}\zeta(3)^{1/3}
\left(15N^{1/6}\frac{a_s}{a_h}\right)^{2/5}.
\end{eqnarray}
Here, $a_h=\sqrt{\hbar/m(\omega_z\omega_r^2)^{1/3}}$ denotes the
average harmonic oscillator length scale. At temperatures above
the $T_C$, $\mu$ can be determined from
$\mathrm{Li}_3(z)=\zeta(3)/x^3$, where $Li_3(.)$ is the
third-order polylogarithm function. The semi-ideal model has a wide-range of validity
in representing density distribution of a trapped Bose gas at finite temperature provided
that $s<0.4$\cite{naraschewski}. At the same time the interactions are assumed to be strong
enough to ensure $\mu\gg \hbar\omega_{r,z}$ so that kinetic energy of the condensate can be neglected
according to the Thomas-Fermi approximation. In typical slow-light experiments in cold atomic gases,
$s$ remains within these limits. We shall use the semi-ideal model given density profile $\rho$
in our investigations of propagation and dispersion of slow pulses and present our explicit results in
the next section.
\section{Results and discussions}

\begin{figure}[htbp]
\centering
\includegraphics[width=8.3cm]{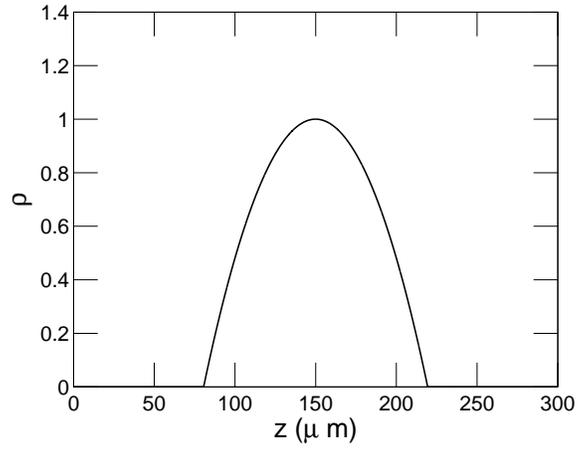}
\caption{Axial density profile of $^{23}$Na the Bose-Einstein
condensate of $N=8.3\times10^6$ atoms at $T=43$ nK. $\rho$ is scaled by the peak density. Trap
parameters are chosen to be $\omega_{r}=2\pi\times69$ Hz and
$\omega_{z}=2\pi\times21$ Hz. tarhanF2.eps.}
\label{fig2}
\end{figure}

In our numerical calculations of the theoretical results presented
in the preceding section, we shall specifically consider a gas of
$N=8.3\times10^6$ $^{23}$Na atoms for which $M=23$ amu, $\lambda_0
= 589$ nm, $\gamma = 2\pi\times 10.01$ MHz, $\Gamma_3=0.5\gamma$,
$\Gamma_2=2\pi\times 10^3$ Hz, and $a_{s}=2.75$ nm. For the
parameters of the trapping potential, we take
$\omega_{r}=2\pi\times69$ Hz and $\omega_{z}=2\pi\times21$ Hz as
in Ref.\cite{slowlight-exp1}. The coupling field Rabi frequency is
taken to be $\Omega_c=0.56\gamma$\cite{slowlight-exp1}. Critical
temperature for Bose-Einstein condensation of such a gas is found
to be $T_C=424$ nK. We illustrate the spatial density profile of
the condensate in the axial ($z$) direction at $T=43$ nK in
Fig.\ref{fig2}.
%
%
%
%

Due to spatially inhomogeneous density, the linear dielectric
susceptibility would also be spatially inhomogeneous. In this
case, group velocity defined by Eq.(\ref{vg}) would have a local
character. Under EIT conditions, when the light pulse enters the
condensate from the thermal component of the gas, its group speed
exhibits a dramatic slowing down as shown in Fig.\ref{fig3}. Here
we consider resonant probe pulse with $\Delta=0$. Within the
condensate region, at such low temperatures, the group velocity
remains roughly at the same ultraslow value. The pulse rapidly
accelerates to high speeds when it leaves the condensate at the
interface to thermal component. Before and after the thermal
component, the pulse is assumed to be propagating in vacuum. In
practice, group velocity is measured in terms of time delay of the
pulse with respect to a reference pulse propagating in vacuum over
the same distance with the gas. To make comparisons with this
operational definition of the group velocity, one needs to make
careful spatial averaging of the theoretical group
velocity\cite{morigi}.
%
%
%
%

The other coefficients in the wave equation are also locally
defined. The coefficient of absorption and the second-order
dispersion coefficient are shown in Fig.\ref{fig4} and in
Fig.\ref{fig5}, respectively.
%
%
%
%
Under EIT conditions, imaginary part $\chi^{\prime\prime}$ of
$\chi$ is small, despite the large condensate density. For large
$\Omega_c$ we have
$\chi^{\prime\prime}\sim\rho\Gamma_2/\Omega_c^2$. For short pulse
propagation we shall see that the major source of reduction in the
peak of the pulse would be the temporal broadening due to
dispersion. The absorption over the small size of the BEC is
negligibly small. Taking $\alpha= 10^3\,$m$^{-1}$ and
$L=100\,\mu$m, we can roughly estimate attenuation of the pulse by
$\exp{(-\alpha L)}$, which yields about $90\%$ transmission. The
loss term can be made smaller by using larger Rabi frequencies for
the coupling control field. The positive sign of $\alpha$ reflects
its effect on the pulse as the decrease of the pulse intensity.
The broadening of the pulse on the other hand is independent of
the sign of $b_2$, as can be deduced from Eq.\ref{eq:pulsewidth}.

%
%
%
%
The region over which dispersive effects are most strong is a
small region near the center of the cloud as can be seen in
Fig.\ref{fig5}. The size of this region can be determined by the
width $L$ of the density distribution, since $b_2$ is proportional
to $\rho$. It is quite smaller than the classical boundary
$\sqrt{2\mu/m\omega_z^2}$. As an example, $L\approx 26\,\mu$m in
Fig.\ref{fig2}, while the condensate is extended over a size of
$\sim 140\,\mu$m. In such a small region, spatial variations of
$\rho$ can be ignored so that
$\rho\approx\rho(r=0,z=0)\equiv\rho_0$. When a pulse of width
$\tau_0$ enters into this strongly dispersive zone, its final
width would approximately broaden to
$\tau(L)=\tau_0\sqrt{1+\kappa^2(\rho_0 L)^2/\tau_0^4}$, where
$\kappa=3\Gamma_3\gamma\lambda^2/\pi^2\Omega_c^4$. Broadening of
the pulse depends on the density profile of the medium through its
characteristic parameters, its peak value $\rho_0$, and its width
$L$. For small dispersion, change in the pulse width would be
proportional to $(\rho_0 L)^2$. For the given $\rho$ within the
semi-ideal model, $L$ demonstrates the well-known temperature
variation\cite{giorgini} as shown in Fig.\ref{fig6}. At a given
temperature, $L$ has been evaluated by Eq.\ref{eq:axial_size}. At
very high temperatures, $T \gg T_C$, classical thermal radius
would be $\sim\sqrt{T}$ according to the equipartition theorem. In
the temperature range of Fig.\ref{fig6}, beyond $T_C$, $L$ is
changing linearly with $T$. It might be illuminating to rewrite
Eq.\ref{eq:axial_size} as $L=\sqrt{(4\pi/N)(\langle
z^2\rangle_C+\langle z^2\rangle_T)}$. Here, $\langle
z^2\rangle_{C,T}$ stands for a mean square axial distance
evaluated using either condensate (C) or thermal density (T)
distribution. The sharp drop of the width of the distribution just
below $T_c$ is due to the emergence of condensed component with
relatively high density about the center of the cloud. As $T$
becomes lower, thermal component shrinks, while the condensate
part expands slowly about the center of the cloud. $L$ eventually
saturates down to Thomas-Fermi width at zero temperature. The
thermal behavior of the peak density is also
well-known\cite{naraschewski,giorgini}. At high temperatures it
shows $\sim 1/T^{3/2}$ decay consistent with classical gas
behavior. Just below $T_C$, $\rho_0$ grows rapidly due to emerging
relatively dense condensate about the center of the cloud. At
extremely low temperatures, $\rho_0$ saturates to Thomas-Fermi
density.

Fig.\ref{fig7} describes dependence of broadening on temperature
and interatomic interaction. At extremely low temperatures, $L$
and $\rho_0$ are determined by the condensate part and they change
slowly with $T$. For increasing $T$, thermal part expands beyond
the boundaries of the condensate, so that $L$ will be contributed
significantly by the thermal part. Except at temperatures close to
$T_C$, $\rho_0$ is dominated by the condensate part, which
exhibits slow dependence on $T$ in comparison to that of $L$. As a
result, the product $\rho_0 L$ grows, following the expansion of
thermal component, up to a point just below $T_C$, where $\rho_0$
dramatically drops to dilute thermal density value due to rapid
disappearance of the dense condensate component. After that peak
point, $\rho_0$ shows faster variation with $T$ relative to that
of $L$. Hence, the shape of broadening curve is characterized by
the thermal behavior of $\rho_0$. At temperatures higher than
$T_C$, broadening decreases with temperature. At such high
temperatures, $L$ grows slowly (almost linear) while $\rho_0$
decreases like $1/T^{3/2}$. Thus their product becomes a
decreasing function with $T$. Summarizing, the broadening is
determined by the density profile through its characteristic
product $\rho_0 L$. As $\rho_0$ and $L$ exhibit competing thermal
behaviors, a peak arises just before the $T_C$, a signature of
emerging condensate being a strongly dispersive region. Before the
peak point, it is $L$ which shapes the temperature dependence of
the broadening. After the peak point, broadening behaves analogous
to the thermal behavior of $\rho_0$.

Earlier studies indicate that group velocity depends on the
strength of atomic interactions within the condensate which is
characterized by the s-wave scattering length $a_s$. As the $a_s$
grow longer, the group velocity of the pulse within the BEC
becomes faster\cite{Mustecaplioglu}. In order to discuss the
influence of atomic interactions on broadening of the pulse, we
first note that below $T_C$ the atomic energies are so low that,
there is only $a_s$, s-wave scattering length, to characterize the
strength of interactions within the condensate component. At
higher temperatures beyond $T_C$, the interactions has no effect
within the semi-ideal model where thermal gas is dilute and
non-interacting. In Fig.\ref{fig7}, we see that as the $a_s$
increases, the broadening decreases when $T<T_C$. According to
semi-ideal model density profile, we get $\rho_0=\mu/U$ at
$T<T_C$, from which we conclude $\rho_0\sim 1/a_s^{3/5}$. This is
intuitively expected, as the increase of repulsive atomic
interactions results in lower densities. The width of the atomic
cloud would slightly increase with $a_s$. The increase of the size
of the condensate can be estimated from the Thomas-Fermi axial
radius given by $\sqrt{2\mu/m\omega_z^2}\sim a_s^{1/5}$. Thus, we
see that the increase of $L$ would be slower than the decrease of
the $\rho_0$, so that $\rho_0 L$ and hence the broadening would
reduce with $a_s$. This conclusion remains the same at higher
temperatures where the thermal length scale becomes dominant in
determining $L$, and it is not influenced by $a_s$. Thus,
broadening decreases due to decrease of $\rho_0$.  Beyond $T_C$,
the density profile is semi-classically determined and is
independent of $a_s$. The semi-ideal model density profile becomes
not as good description of the condensate when the interactions
characterized by the scaling parameter $s$ becomes larger than
$0.4$, corresponding to $a_s\sim 7$ nm. For $^{23}$Na cloud,
$a_s=2.75$ nm. The higher $a_s$ may be achieved by utilizing
Feshbach resonance technique\cite{fedichev,bohn,inouye} or by
considering alternative atoms.
%
%
%

Our approach in determining the pulse broadening using the
effective medium width $L$ , as we will show below, gives
qualitatively accurate behavior of the broadening but
overestimates it. In an actual pulse propagation, a pulse of width
$\tau_0$ would be prepared far from the gaseous medium. Before
reaching to the most dispersive zone about the center of the
cloud, the pulse would pass through the thermal cloud and then through the
edges of the condensate. Due to the inhomogeneous density profile,
when the pulse arrives to the center of the cloud, it would
already be much broader than $\tau_0$. As a result it would suffer
less from dispersion than the one predicted above. Despite the
lack of actual input value $\tau_0$ for our analysis above, the
overall qualitative behavior is expected to be accurate. To test
this expectation and to justify our effective central dispersive zone
treatment agree qualitatively with the exact pulse behavior, we now investigate
the variation of the pulse width with distance of propagation,
taking into account the full spatially inhomogeneous density
profile of the gaseous medium. We assume a Gaussian pulse of the
form $\exp{(-\beta (t-t_0)^2)}$ at initial time $t_0$, , where
$\sqrt{\beta}$ is the pulse width, and propagate it numerically. A
dimensionless form of Eq.(\ref{eq:pulse}) is solved via finite
difference Crank-Nicholson space marching scheme. The
Crank-Nicholson scheme is less stable but more accurate than the
fully implicit method; it takes the average between the implicit
and the explicit schemes\cite{Garcia}. We use forward difference
scheme for the position and central difference scheme for the
time. Discrete equations in matrix form are solved using Thomas
algorithm\cite{Garcia} which is analogous to a fast Gaussian
elimination method for tridiagonal matrices.

We have performed extensive numerical simulations at various $T$
and confirmed that despite being an overestimation, the
qualitative behavior predicted in Fig.\ref{fig7} correctly
describes the results obtained by the exact solution. We present
typical results of our simulations in Figs.\ref{fig8}-\ref{fig11}.
For a nanosecond pulse, at $T=43\,$nK, broadening is $\sim
3\,\mu$s which can be seen in Fig.\ref{fig9}. This is much smaller
than the result found by the uniform density calculation, which
gives $\sim 100\,\mu$s at the same temperature in Fig.\ref{fig7}.
More moderate broadening is
obtained for a pulse of width $0.1\,\mu$s and shown in
Fig.\ref{fig10}.
%
%
%
%
%
%
As a rule of thumb, we assume that  broadening is not
significantly large if it is less than a factor of $2$. A
microsecond pulse broadens by a factor of $\sim 1.7$ as can be
seen in Fig.\ref{fig11}. For longer pulses therefore we conclude
that broadening is not significant.

In early ultraslow light experiments microsecond pulses were
used\cite{slowlight-exp1}. Previous theoretical models ignore
dispersion effects in the pulse propagation. While they can be
used to describe observed group velocities, they cannot explain
the transmitted pulse shape adequately. High-order dispersion,
absorption, transverse diffraction, and off-resonant transitions
to upper levels not included in the three-level model cause
imperfect transition, and result in reduction of the transmitted
peak intensity. We have already argued that absorption, though
still present even under EIT conditions, makes a little effect
over the small BEC size. Transverse diffraction due to paraxial
pulse propagation is also not too strong\cite{Mustecaplioglu}. To
characterize the contribution of second-order dispersion on the
reduction of the peak intensity, propagation of a microsecond
pulse is simulated and presented in Fig.\ref{fig11}. This result
indicates that in addition to off-resonant transitions to other
levels, broadening of the pulse also contributes to the observed
drop of the transmitted peak intensity.
%
%
%
%

While microsecond pulses are good to minimize dispersive effects,
shorter pulses may be more attractive for optical information
processing. The question we would like to now address is if such
shorter pulses can still be stored as efficiently via EIT scheme
in atomic ultracold gases. We may characterize coherent optical
information storage capacity through a parameter given by
$C=L/2v_g\tau(L)$. It measures number of probe pulses
simultaneously present in BEC without significant overlap. The
storage time is taken to be the delay time of the pulses,
$t_s=L/v_g$. We choose an ideal repetition rate of the pulse train
to be $1/2\tau(L)$. The pulses are assumed to be stored in the
central region of the cloud, whose length $L$ is given by
Eq.\ref{eq:axial_size}. It is the same with the effective central
dispersive zone considered in the analytical broadening
calculation. In this region $\rho\approx\rho_0$ is approximately
uniform at its peak value, leading to spatially homogeneous $v_g$,
that can be seen in Fig.\ref{fig3}. More explicitly, $C$ becomes
\begin{eqnarray}
C=\frac{L}{2\tau_0\sqrt{4\pi^2\Omega_c^4/9\lambda^4\gamma^2\rho^2
+4L^2\Gamma_3^2/\pi^2\tau_0^4\Omega_c^4}}.
\end{eqnarray}
The second term under the square root in the denominator
represents the second-order dispersive contribution. When
dispersion is not too strong, $C$ could be made higher by
increasing $\rho$. On the other hand, it is not very simple task
to increase $\rho$ for BEC. Besides, it implicitly and indirectly
affects the dispersive term via $L$. For a given $\tau_0$, more
explicit and direct control parameter would be $\Omega_c$. $C$
exhibits a non-trivial dependence on $\Omega_c$, which is
illustrated in Fig.\ref{fig12}. We observe that for given atomic
cloud, and $\tau_0$, there is an optimum choice of $\Omega_c$.
Fig.\ref{fig12} also compares $C$ for different $\tau_0$. We
deduce that, for a wide range of $\Omega_c$ it is indeed possible
to inject more pulses by using shorter $\tau_0$. However, reducing
$\tau_0$ from $1\,\mu$s to $0.01\,\mu$s, one cannot hope to
increase the capacity $100$ times more. Instead, due to dispersive
effects, only by using a critical coupling laser Rabi frequency,
storage capacity can be made maximum. The maximum achievable
capacity $C_{max}$ is independent of $\tau_0$, and is found to be
\begin{eqnarray}
C_{max}=\sqrt{\frac{3\gamma\lambda^2L\rho}{32\Gamma_3}},
\end{eqnarray}
which is evaluated at the critical coupling Rabi frequency,
$\Omega_{c0}$, given by
\begin{eqnarray}
\Omega_{c0}=\left(\frac{3\Gamma_3\lambda^2\gamma\rho L}{\pi^2\tau_0^2}\right)^{1/4}.
\end{eqnarray}
These results suggest that shorter pulses can be stored in the
cloud as efficiently as microsecond pulses. The storage time
$t_{s0}$ when $\Omega_c=\Omega_{c0}$ becomes
\begin{eqnarray}
t_{s0}=\tau_0\left(\frac{\sqrt{3}\lambda\sqrt{\gamma}}{2\sqrt{\Gamma_3}}\right)
\left(\sqrt{\rho L}\right).
\end{eqnarray}

We note that analytical result of $C$ underestimates the actual
value of it, as $\tau(L)$ is overestimated. Due to spatial
inhomogeneity, $\tau_0$, pulse width just before the central zone,
would be larger than the width of the original pulse prepared away
from the atomic cloud, so that $\tau(L)$ would be smaller. $v_g$
would remain the same under any pulse width change. As a result,
actual $C$ would be higher than the one predicted above. The
analytical expressions are mainly good for predicting qualitative
behavior of $C$ with $\Omega_C$, and providing a lower limit for
its value.
%
%
%
%
%
%
\section{Conclusion}

We have examined dispersive effects on short pulse propagation
through a semi-ideal BEC under EIT scheme. We have found that third and
higher order dispersion coefficients are negligibly small for
current experimental systems. Distortion of the shape of
a short pulse in the atomic cloud would be a temporal broadening arising from second-order
dispersion per se. For pulses of widths greater than microseconds we
have found that second order dispersion is also negligible. For microsecond pulses
the broadening is about $1.7$ of the initial width. As a rule of
thumb, when the pulse length is broadened by less than a factor of $2$,
the pulse can be considered to be not significantly distorted.

Second order dispersion of the atomic cloud depends on its
density profile, which is determined mainly by temperature and
atomic interactions. For nanosecond pulses, we
have presented the dependence of the broadening on the
temperature and interatomic scattering. The broadening is found to
be making a peak just below the critical temperature $T_C$. As
the scattering length increases, the broadening decreases at all
temperatures, resulting in a lower peak.

Furthermore, we have numerically examined the effect of spatial
variations of the atomic cloud. Our results indicate that due to
spatial inhomogeneity the broadening is less than that predicted
by the calculations based upon effective central dispersive zone.
As the pulse is already broadened relative to its original width,
when it arrives at the central zone, it suffers less from the high
dispersive properties of this region. Qualitatively, we have found
effective central zone estimations describe the behavior of the
broadening very well. Our numerical simulations showed that the
transmitted pulse shape in the ultraslow light experiments starts
to get distorted for pulses of widths shorter than or equal to
microsecond.

Finally we have shown that despite the dispersive effects it is
still possible to achieve significant increase of bit storage
capacity by carefully choosing a critical coupling field Rabi
frequency. Though shorter pulses allow for higher capacity for a
wide range of Rabi frequencies of the coupling field, the maximum
achievable capacity is independent of the initial pulse width of
the probe pulse and determined by the properties of the atomic
cloud only.

Our results may be used for measuring $T_C$ and $a_s$ of the BEC medium
by comparing the group speed and the broadening of the probe pulse with a reference
pulse propagating in the absence of medium.
One may also exploit Feshbach resonances to tune $a_s$ to
control broadening. Further uses of our results may be found in pulse shape
engineering, frequency filtering, and enhancing capacity of dynamic optical memories.

\acknowledgements

We thank S. Sefi for help in numerical computations and
N. Postac\i{}o\~{g}lu for discussions.
\"O.E.M. acknowledges support from T\"UBA-GEB\.{I}P Award. This
work was partially supported by Istanbul Technical University Foundation
(ITU BAP) under Project No. 31192.

\newpage

\section*{List of Figure Captions}

Fig. 1. Schematic energy level diagram of a
three-level atom interacting with two laser
beams in $\Lambda$ configuration.

\noindent Fig. 2. Axial density profile of $^{23}$Na the Bose-Einstein
condensate of $N=8.3\times10^6$ atoms at $T=43$ nK. $\rho$ is scaled by the peak density. Trap
parameters are chosen to be $\omega_{r}=2\pi\times69$ Hz and
$\omega_{z}=2\pi\times21$ Hz.

\noindent Fig. 3. Spatial dependence of the local group velocity of a
resonant probe pulse along the $z$-axis, with $\Delta=0$,
propagating through a $^{23}$Na the Bose-Einstein condensate of
$N=8.3\times10^6$ atoms at $T=43$ nK under EIT scheme. The
parameters used are $M=23$ amu, $a_{s}=2.75$ nm, $\lambda_0 = 589$
nm, $\gamma = 2\pi\times 10.01$ MHz, $\Gamma_3=0.5\gamma$, $\Omega_c=0.56\gamma$,
$\Gamma_2=2\pi\times 10^3$ Hz.

\begin{figure}[htbp]
\centering
\includegraphics[width=8.3cm]{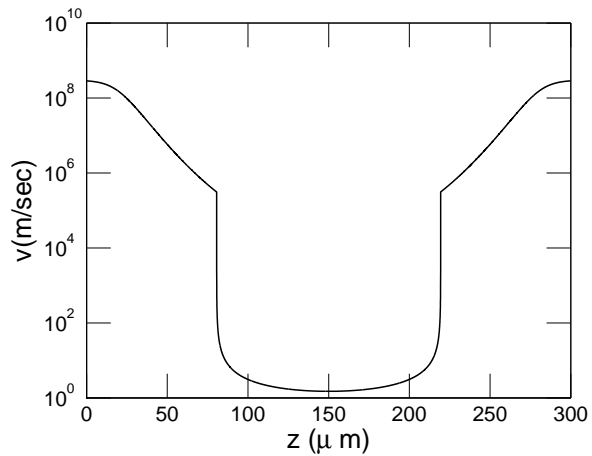}
\caption{Spatial dependence of the local group velocity of a
resonant probe pulse along the $z$-axis, with $\Delta=0$,
propagating through a $^{23}$Na the Bose-Einstein condensate of
$N=8.3\times10^6$ atoms at $T=43$ nK under EIT scheme. The
parameters used are $M=23$ amu, $a_{s}=2.75$ nm, $\lambda_0 = 589$
nm, $\gamma = 2\pi\times 10.01$ MHz, $\Gamma_3=0.5\gamma$, $\Omega_c=0.56\gamma$,
$\Gamma_2=2\pi\times 10^3$ Hz. tarhanF3.eps.}
\label{fig3}
\end{figure}

\begin{figure}[htbp]
\centering
\includegraphics[width=8.3cm]{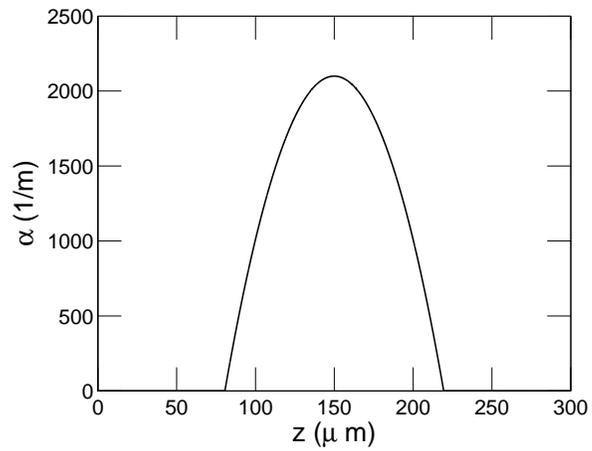}
\caption{Position dependence of the absorption coefficient
$\alpha$ along the $z$-axis. The parameters are the same with
those of Fig.\ref{fig3}. tarhanF4.eps.}
\label{fig4}
\end{figure}

\begin{figure}[htbp]

\centering
\includegraphics[width=8.3cm]{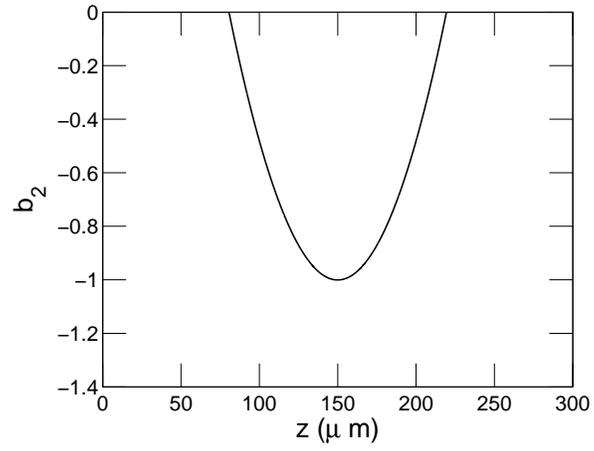}
\caption{Axial spatial profile of the second-order dispersion
coefficient $b_2$ for the parameters same with those of Fig.\ref{fig3}.
Vertical axis is scaled by the peak value of $b_2$
which is $\sim 0.108$ s$^2/$nm. tarhanF5.eps.}
\label{fig5}
\end{figure}

\begin{figure}[htbp]
\centering
\includegraphics[width=8.3cm]{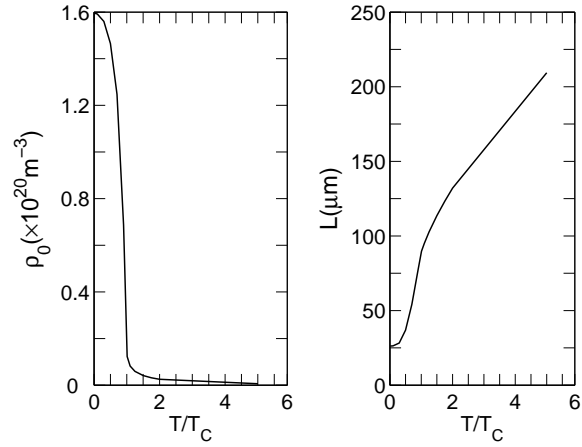}
\caption{Thermal behavior of the peak $\rho_0$ and the effective
axial length $L$ of the atomic density distribution, given by the
semi-ideal model. The parameters are the same with those of
Fig.\ref{fig3}. tarhanF6.eps.} \label{fig6}
\end{figure}

\begin{figure}[htbp]
\centering
\includegraphics[width=8.3cm]{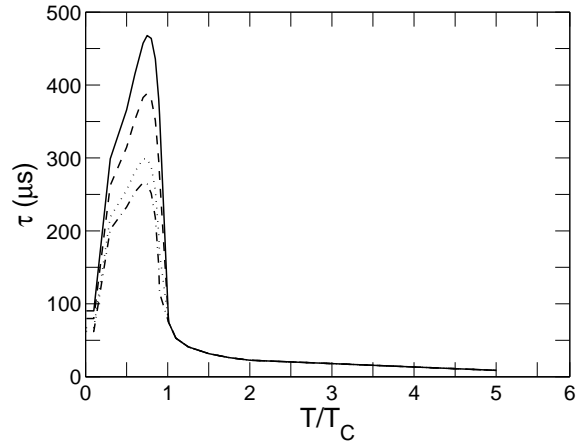}
\caption{Thermal behavior of the broadening of a nanosecond pulse
obtained analytically by an effective dispersive zone treatment.
The solid, dashed, dotted and dot-dashed lines are for $a_s=2.75,
3.75, 5.75$ nm, and $a_s=7$ nm, respectively. The other parameters
are the same with those of Fig.\ref{fig3}. tarhanF7.eps.}
\label{fig7}
\end{figure}

\begin{figure}[htbp]
\centering
\includegraphics[width=8.3cm]{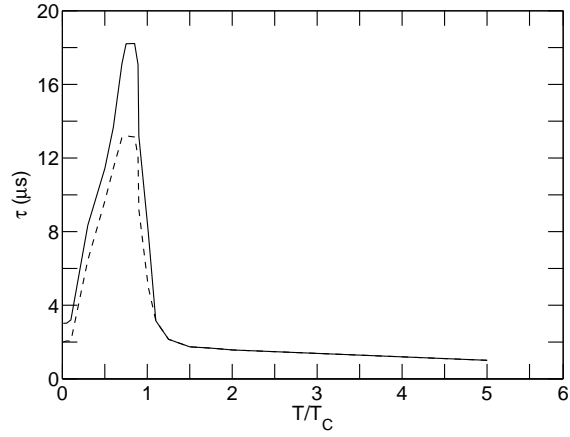}
\caption{Thermal behavior of the broadening of a nanosecond pulse
determined by numerical simulations. The solid and dashed lines are for
$a_s=2.75$ nm and $a_s=7$ nm, respectively. The other parameters are the same
with those of Fig.\ref{fig3}. tarhanF8.eps.} \label{fig8}
\end{figure}

\begin{figure}[htbp]
\centering
\includegraphics[width=8.3cm]{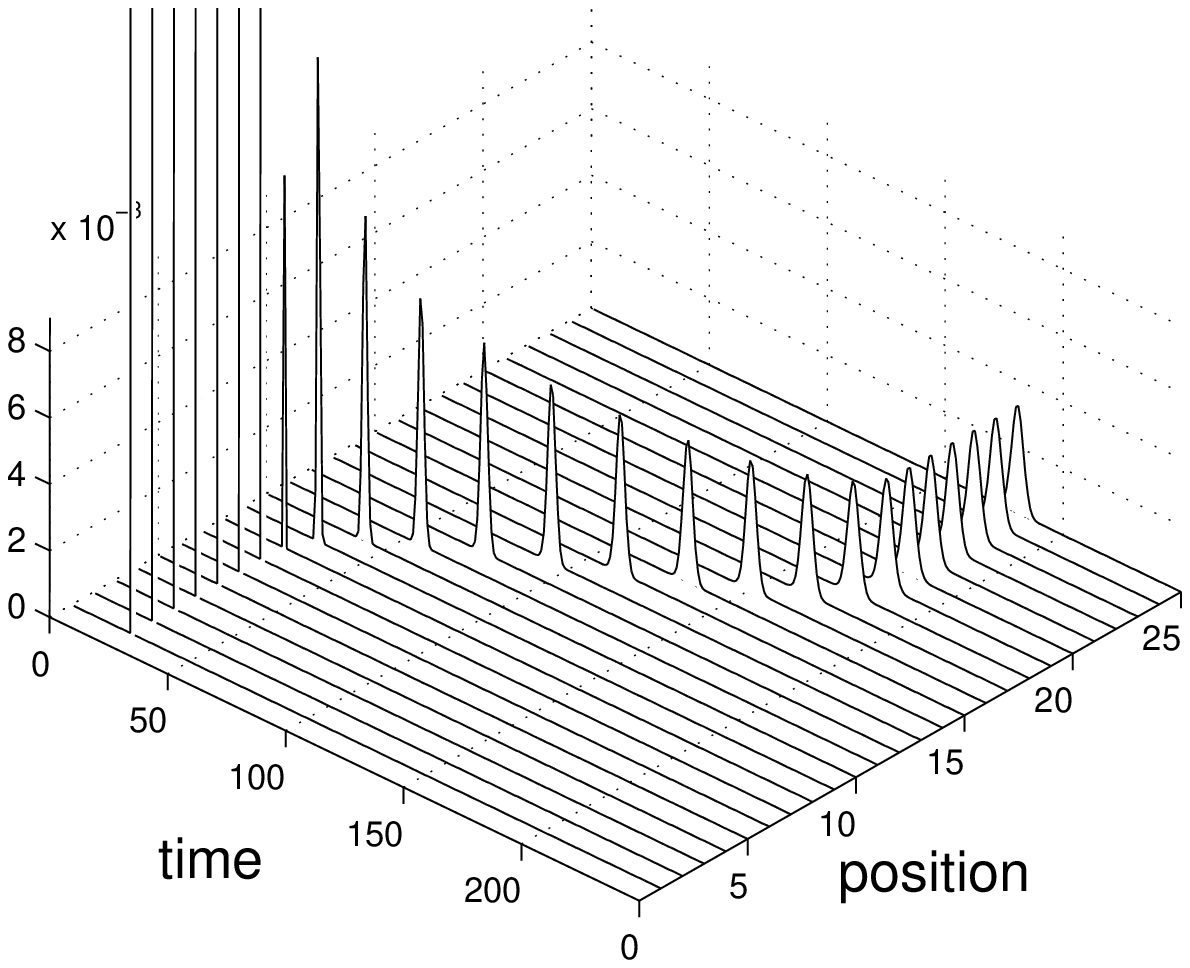}
\caption{Propagation of a nanosecond pulse through the interacting
BEC. Time ($t$) is scaled by $0.4\,\mu$s. and position ($z$) is by
$10\,\mu$m. The parameters are the same with those of
Fig.\ref{fig3}. tarhanF9.eps.} \label{fig9}
\end{figure}

\begin{figure}[htbp]
\centering
\includegraphics[width=8.3cm]{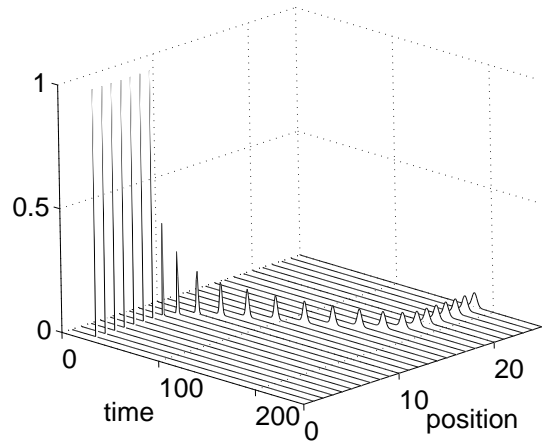}
\caption{Propagation of a pulse of width $0.1\,\mu$s through the
interacting BEC. Time ($t$) is scaled by $0.4\,\mu$s. and position
($z$) is by $10\,\mu$m. The parameters are the same with those of
Fig.\ref{fig3}. tarhanF10.eps.} \label{fig10}
\end{figure}

\begin{figure}[htbp]
\centering
\includegraphics[width=8.3cm]{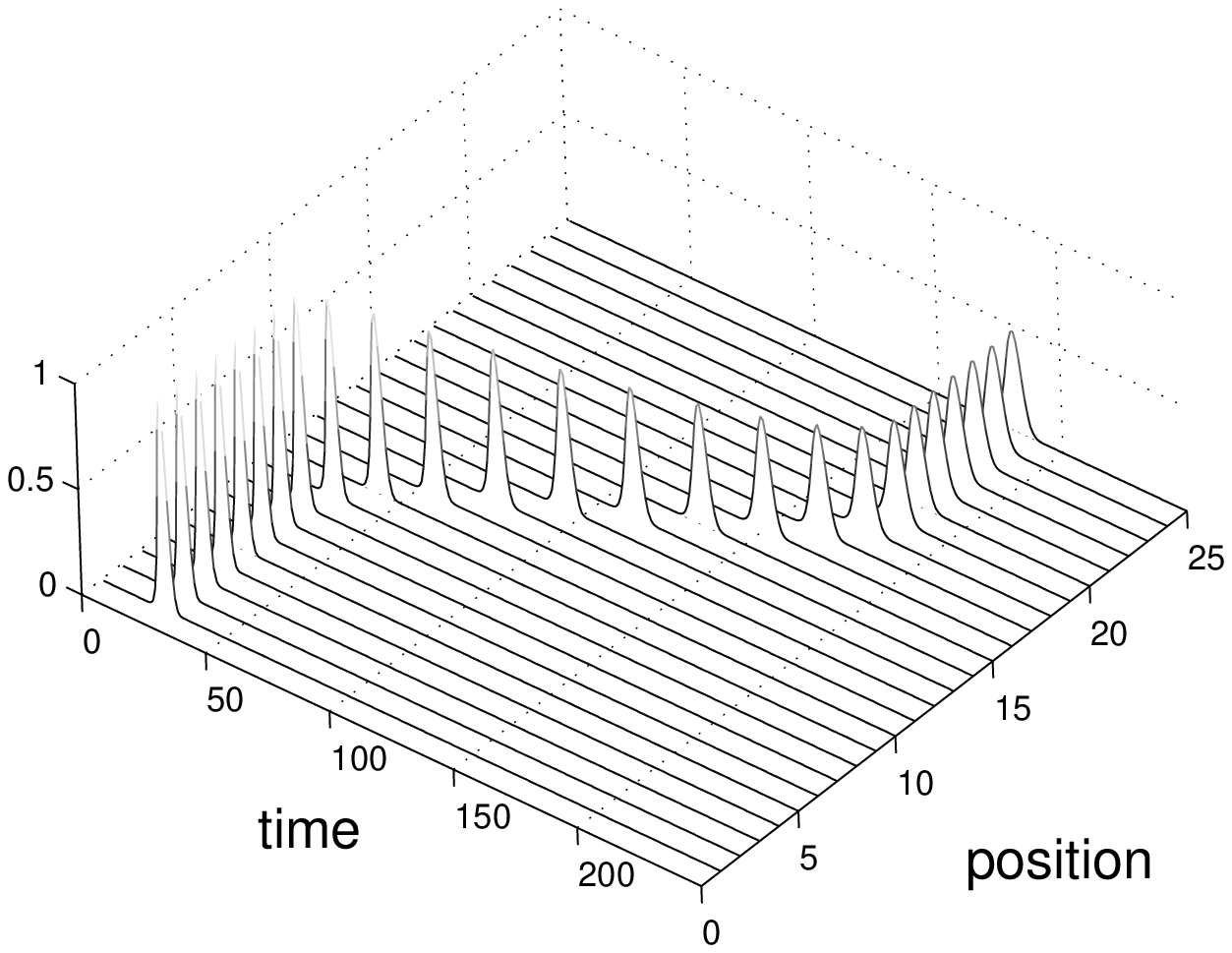}
\caption{Propagation of a microsecond pulse through the
interacting BEC. Time ($t$) is scaled by $0.4\,\mu$s. and position
($z$) is by $10\,\mu$m. The parameters are the same with those of
Fig.\ref{fig3}. tarhanF11.eps.} \label{fig11}
\end{figure}

\begin{figure}[htbp]
\centering
\includegraphics[width=8.3cm]{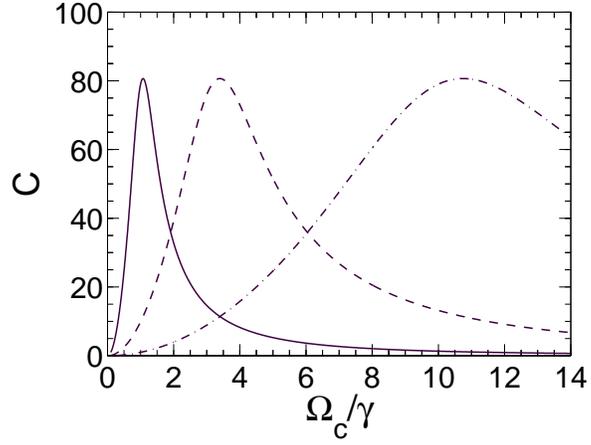}
\caption{Coherent optical information storage capacity $C$
(dimensionless) as a function of coupling field Rabi frequency
$\Omega_c$, scaled by $\gamma$. Parameters used are $L=100\,\mu$m,
$\tau_0=1\,\mu$s (solid line), $\tau_0=0.1\,\mu$s (dashed line),
$\tau_0=0.01\,\mu$s (dot-dashed line), $\rho=10^{21}\,$m$^{-3}$.
The other parameters are the same with those of Fig.\ref{fig3}.
tarhanF12.eps.} \label{fig12}
\end{figure}

\noindent Fig. 4. Position dependence of the absorption coefficient
$\alpha$ along the $z$-axis. The parameters are the same with
those of Fig.\ref{fig3}.

\noindent Fig. 5. Axial spatial profile of the second-order dispersion
coefficient $b_2$ for the parameters same with those of Fig.\ref{fig3}.
Vertical axis is scaled by the peak value of $b_2$
which is $\sim 0.108$ s$^2/$nm.

\noindent Fig. 6. Thermal behavior of the peak $\rho_0$ and the
effective axial length $L$ of the atomic density distribution,
given by the semi-ideal model. The parameters are the same with
those of Fig.\ref{fig3}.

\noindent Fig. 7. Thermal behavior of the broadening of a
nanosecond pulse obtained analytically by an effective dispersive
zone treatment. The solid, dashed, dotted and dot-dashed lines are
for $a_s=2.75, 3.75, 5.75$ nm, and $a_s=7$ nm, respectively. The
other parameters are the same with those of Fig.\ref{fig3}.

\noindent Fig. 8. Thermal behavior of the broadening of a
nanosecond pulse determined by numerical simulations. The solid
and dashed lines are for $a_s=2.75$ nm and $a_s=7$ nm,
respectively. The other parameters are the same with those of
Fig.\ref{fig3}.

\noindent Fig. 9. Propagation of a nanosecond pulse through the
interacting BEC. Time ($t$) is scaled by $0.4\,\mu$s. and position
($z$) is by $10\,\mu$m. The parameters are the same with those of
Fig.\ref{fig3}.

\noindent Fig. 10. Propagation of a pulse of width $0.1\,\mu$s
through the interacting BEC. Time ($t$) is scaled by $0.4\,\mu$s.
and position ($z$) is by $10\,\mu$m. The parameters are the same
with those of Fig.\ref{fig3}.

\noindent Fig. 11. Propagation of a microsecond pulse through the
interacting BEC. Time ($t$) is scaled by $0.4\,\mu$s. and position
($z$) is by $10\,\mu$m. The parameters are the same with those of
Fig.\ref{fig3}.

\noindent Fig. 12. Coherent optical information storage capacity
$C$ (dimensionless) as a function of coupling field Rabi frequency
$\Omega_c$, scaled by $\gamma$. Parameters used are $L=100\,\mu$m,
$\tau_0=1\,\mu$s (solid line), $\tau_0=0.1\,\mu$s (dashed line),
$\tau_0=0.01\,\mu$s (dot-dashed line), $\rho=10^{21}\,$m$^{-3}$.
The other parameters are the same with those of Fig.\ref{fig3}.

\newpage
\newpage
\newpage
\newpage
\newpage
\newpage
\newpage
\newpage
\newpage
\newpage
\newpage
\newpage
\end{document}